# Structure and Phase Transitions of Alkyl Chains on Mica

by

Hendrik Heinz, Hein J. Castelijns[*] and Ulrich W. Suter[§]

Department of Materials, Institute of Polymers, ETH, CH-8092 Zürich, Switzerland

[*] Present address:

Department of Applied Physics, Technical University of Eindhoven, P.O. Box 513, NL-5600

MB Eindhoven, The Netherlands

[§] corresponding author: uwsuter@eth.ch

# Abstract

We use molecular dynamics as a tool to understand the structure and phase transitions [Osman et. al. *J. Phys. Chem. B* 2000, *104*, 4433-4439; Osman et. al. *J. Phys. Chem. B* 2002, *106*, 653-662] in alkylammonium micas. The consistent force field 91 is extended for accurate simulation of mica and related minerals. We investigate mica sheets with 12 octadecyltrimethylammonium ($C_{18}$) ions or 12 dioctadecyldimethylammonium ($2C_{18}$) ions, respectively, as single and layered structures at different temperatures with periodicity in the *xy* plane by NVT dynamics. The alkyl-lammonium ions reside preferably above the cavities in the mica surface with an aluminum-rich boundary. The nitrogen atoms are 380 to 390 pm distant to the superficial silicon-aluminum plane. With increasing temperature, rearrangements of $C_{18}$ ions on the mica surface are found, while $2C_{18}$ ions remain tethered due to geometric restraints. We present basal-plane spacings in the duplicate structures, tilt angles of the alkyl chains, and *gauche-trans* ratios to analyze the chain conformation. Agreement with experimental data, where available, is quantitative. In $C_{18}$-mica with less than 100 % alkali-ion exchange, the disordered $C_{18}$ rods in the island structures [Hayes and Schwartz *Langmuir* 1998*, 14,* 5913-5917] break at 40 °C. At 60 °C, the headgroups of the $C_{18}$ alkyl chains rearrange on the mica surface and the broken chain backbones assume a coil-like structure. The $C_{18}$-mica obtained on fast cooling of this phase is metastable due to slow reverse rearrangements of the headgroups. In $2C_{18}$-mica with 70-80 % ion exchange, the alkali ions are interspersed between the alkyl chains, corresponding to a single phase on the surface. The observed phase transition at ~53 °C involves an increase of chain disorder (partial melting) of the $2C_{18}$ ions without significant rearrangements on the mica surface. We propose a geometric parameter $\lambda$ for the saturation of the surface with alkyl chains, which determines the preferred self-assembly pattern, i. e., islands, intermediate, or continuous. $\lambda$ allows the calculation of tilt angles in continuous layers on mica or other surfaces. The thermal decomposition seems to be a Hofmann elimination with mica as a base-template.

# 1. Introduction

Mica was originally a waste product from minery, but extraordinary electrical insulating properties and high thermal stability lead to many applications.[1, 2] Moreover, mica became a cheap filler in joint cement, paints, plastics, and rubber.[3] Mica addition increases stiffness, ductility, and high-heat dimensional stability in plastics, and improves resiliency in rubber. A limiting factor for its utility is, however, that mica flakes may delaminate from nonpolar host materials under shear.

The poor adhesion of mica to nonpolar materials is due to the fact that mica is hydrophilic (see Section 2) and, accordingly, only the interaction in aqueous cement and water-based paints seems favourable. In mainly hydrophobic plastics and rubber, the components remain separable. The interaction with hydrophobic matrices might be significantly improved by changing the polarity of the surface of the mica sheets, for example by exchange of the natural alkali ions against alkylammonium ions of a certain length. Then the mica surface is rendered nonpolar and the interfacial free energy with the adjoining organic polymeric material is strongly reduced. The resulting nanocomposites may obtain greater strength and resiliency.

The properties of such organically modified mica have been the subject of several experimental studies,[4-13] however, a large amount of data on the structure and phase transitions remains unclear.[12, 13] Especially the interaction between the polar mineral surface and the alkylammonium ions has not yet been investigated in detail. Molecular dynamics is a powerful tool for such purpose.[14-22]

In the following section, we discuss some structural details of mica and the modification by ion exchange. In Section 3, we describe the development of our forcefield, which allows accurate simulation of several unit cells of mica, or similar minerals, and attached organic ammonium ions. In Section 4, we present the results from our molecular dynamics simulation concerning the structure of the interface, inclination angles, basal-plane spacings, and conformational analysis of the alkyl chains in neat octadecyltrimethylammonium mica ($C_{18}$-mica) and dioctadecyldimethylammonium mica ($2C_{18}$-mica). Section 5 contains a juxtaposition of the simulation results with the available experimental information. We discuss the structures with less than quantitative ion exchange and explain the observed phase transitions. Thereafter, a basic geometric model is introduced to rationalize the occuring structural patterns. Mechanisms for ion exchange and thermal elimination are suggested. Our paper concludes with a summary in Section 6.

## 2. Mica Structure, Ion Exchange, and Cation Exchange Capacity (CEC)

We consider muscovite $2M_1$, the most abundant of the micas and one of the most stable soil minerals. Ion exchange of the surface alkali cations (see Figure 1) against ammonium ions on treatment with aqueous $(NH_4)_2SO_4$ solutions had been found[23] before the crystal structure of the mineral was formulated.[24] Mica, $K[AlSi_3O_8][AlO(OH)]_4[O_8Si_3Al]K$ (the alkali ions are given as K for simplicity, but are often mixtures, in nature), is a laminated mineral, where each lamella consists of a three-layer sequence (see Figure 1b), as indicated in the formula.

                    

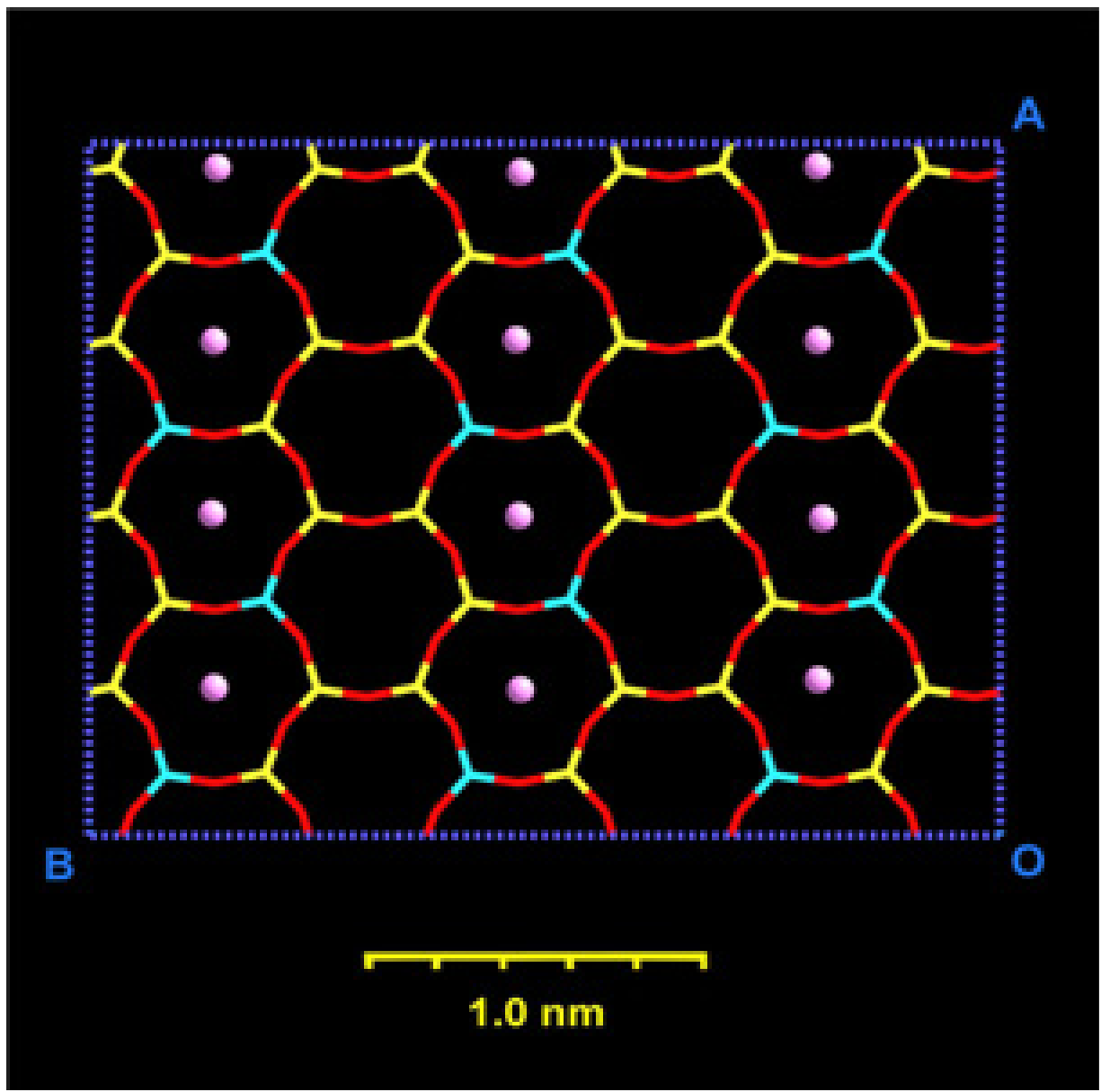

(a)

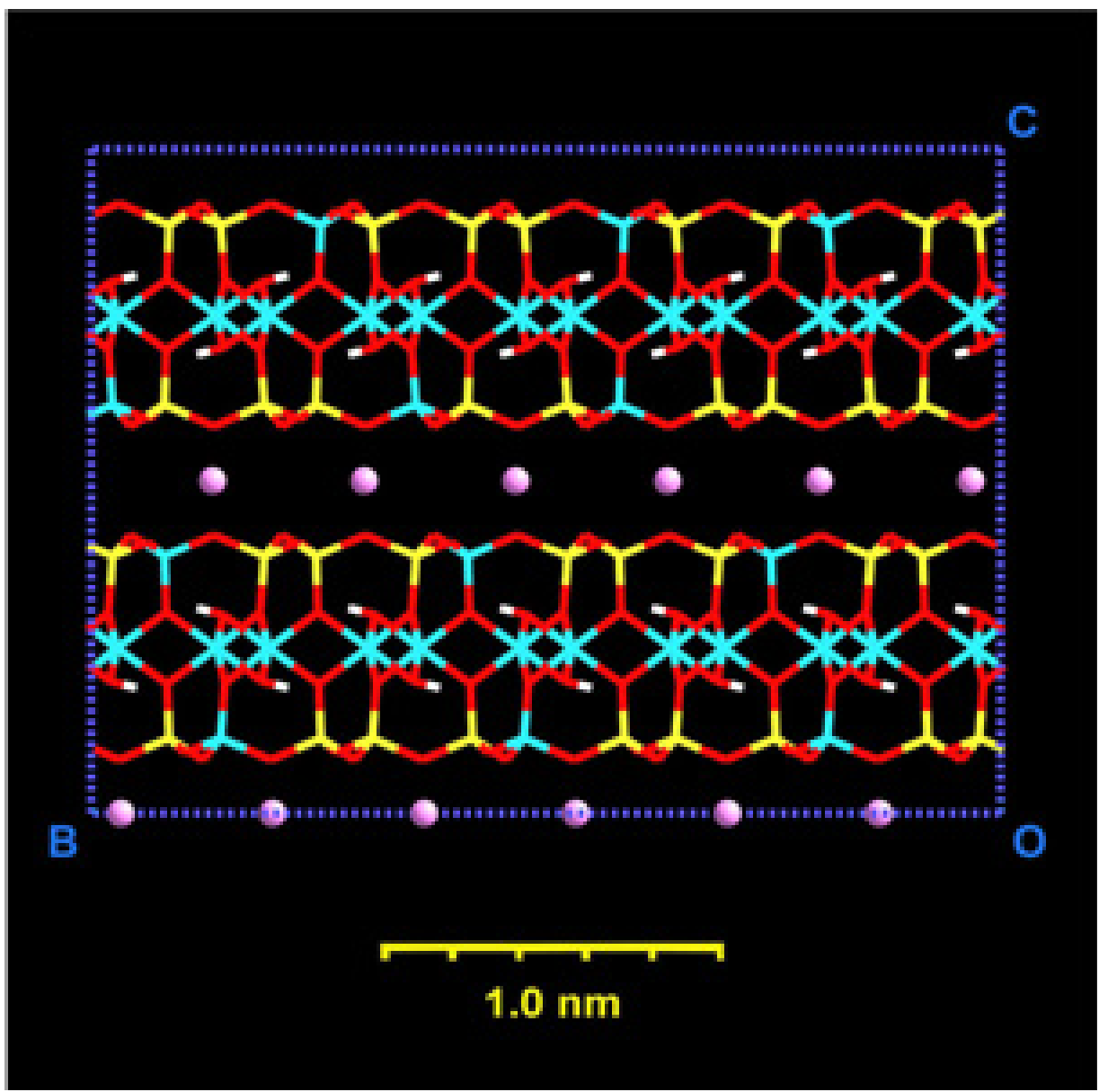

(b)

**Figure 1.** A conglomerate of 4 × 3 × 1 mica unit cells. Al atoms are depicted in blue, Si yellow, O red, H white, and K (or Li) ions in violet. (a) Top view onto a cleaved mica sheet along the *c* direction. Only the upper tetrahedral layer is shown. The Al atoms are distributed in an idealized para-arrangement (ref. 28). The negative charges on oxygen atoms bonded to aluminum are higher than on oxygen atoms bonded only to silicon (ref. 25). The position of the alkali ions corresponds to the energetically most stable arrangement. (b) Side view along the *a* direction. The cavities in the lamellas are fully occupied with alkali ions and stacked on each other. The formation of duplicate mica structures with organic ions can be imagined by intercalation of an amphiphilic bilayer.

## 2.1. Structural Details

(1) There is a tetrahedral layer with the composition of silicon dioxide. Approximately one out of four silicon atoms is replaced by aluminum. (2) The oxygen atoms normal to the surface of the tetrahedral layer are connected to two Al atoms of an inner octahedral layer. The composition can be described as AlO(OH), where we formally count one fraction of the connecting O atoms to the tetrahedral layer and another equal fraction to the octahedral layer. A third fraction of O atoms is only connected to H and two Al atoms (see Figure 1b). (3) On the other side of the octahedral layer, a second tetrahedral aluminosilicate layer is attached, which has the same composition as the previous one. (4) Such lamellas, each consisting of three layers, are stacked, yielding the laminated structure typical for phyllosilicates. The alkali ions, e. g., potassium, are located between the lamellas (see Figure 1).

The aforementioned $Si \rightarrow Al$ substitution in the tetrahedral layers introduces a deficiency of one valence electron, which is offset by the valence electrons from an alkali ion between the lamellas. The balance of partial charges within mica is rather intricate because, generally, a mixture of covalent and electrostatic bonding is present.[25] Accurate modeling of the charge distribution is important for a serviceable simulation.[25]

Moreover, the extent of $Si \rightarrow Al$ substitution is directly related to the available surface area per cation. The ratio usually amounts to $Al_{1.00}Si_{3.00}$ in natural mica,[26-29] but may deviate in less common species down to $Al_{0.75}Si_{3.25}$.[29, 30] The available surface area per alkylammonium ion is determined by this stoichiometry and the percentage of alkali-ion exchange.

## 2.2. Cation Exchange Capacity (CEC)

Although many workers have prepared mica with exchanged cations,[4-13, 23, 31-34] there is confusion about an accurate method to determine the amount of exchangeable cations per mass unit of ground mica.[12, 13, 34] This may be, firstly, because the cation exchange capacity depends on the cations present in the mica. The degree of lamination is lowest for $Li^+$ in the interlayer and magnifies for $Na^+$, $K^+$, $Rb^+$, towards $Cs^+$. Therefore, the cation exchange capacity is highest for Li-mica and diminishes towards Cs-mica.[33] Secondly, in exchange experiments, alkaline cations substitute often to some extent with other than the desired cations, e. g., with $H^+$ at lower pH values or traces of ammonium ions. Thirdly, exchange reactions may not be quantitative.

For example in eq 1,[34]

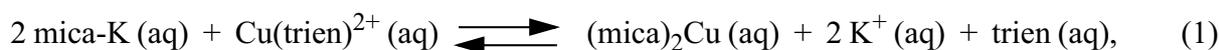

$$2 \text{ mica-K (aq)} + \text{Cu(trien)}^{2+} \text{ (aq)} \rightleftharpoons (\text{mica})_2\text{Cu (aq)} + 2 \text{ K}^+ \text{ (aq)} + \text{trien (aq)}, \qquad (1)$$

where trien designates triethylenetetraamine, potassium release is ~10 % less than with $Cs^+$ as a reaction partner and ~10 % of the released potassium is exchanged against something else than copper.[34] Measuring the $Cu(trien)^{2+}$ adsorption will underestimate the CEC by ~20 %.

In the experimental work on $C_{18}$-mica[12] and $2C_{18}$-mica (see Figure 2) to be used later in this paper,[13] the CEC is known only with limited precision.[32-34] Although the reactive Li-mica was used, a stated extent of 80 % ion exchange might actually be between 70 % and 80 %. This is of some importance in Section 5.

For an accurate estimation of the CEC, we recommend the use of Li-mica and $Cs^+$ ions, which makes the reaction quantitative. The amount of released $Li^+$ is then an accurate measure of the CEC because it is not affected by possible substitution against other cations present.

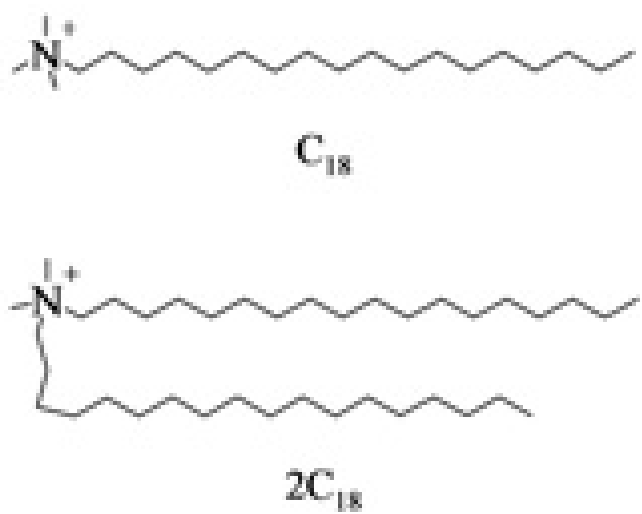

**Figure 2.** Sketch of an octadecyltrimethylammonium ion ($C_{18}$) and a dioctadecyldimethylam-monium ion ($2C_{18}$).

# 3. Force Field Development and Simulation Setup

We need a force field to model both the inorganic mineral and the organic residues. However, modeling of strongly electrostatic interactions makes different demands than modeling of mainly covalent and dispersive interactions. Using conventional energy expressions, proximity to physical reality for the overall system is difficult to achieve.

## 3.1. Force Field Development

The consistent force field 91 (cff91)[35] is specially designed for covalent organic matter and reproduces crystal structures, liquid densities, torsional barriers, and vaporization energies of alkanes well.[36-38] This is mainly related to the physically more realistic, "softer" 9-6 nonbond potential compared to the "harder" 12-6 potential in the consistent valence force field (cvff).[35, 39] (However, spectral shift calculations indicate that an $r^{-4}$ distance dependence of the dispersive energies might be more physical.[38]) The organic part of our system requires a good reproduction of torsional barriers because they are the major force field parameters to determine the degree of disorder in the chains (see Table 1).[40-43] Structural deviations from refined alkane geometries[44] and a known $2C_{18}$ bromide monohydrate structure[45] are only on the order of 1 % in the scaled coordinates.

**Table 1:** The rotational barrier in ethane and the conformational energies in *n*-butane, relative to the *anti*-form (in kcal/mol). Determined from molecular mechanics with angle restraints (ref. 35).

|          | ethane | *n*-butane | | |
|----------|--------|------------|------------|------------|
|          |        | gauche     | part. eclipsed | fully eclipsed |
| Exptl/Lit | 2.8[a] | 0.70[b, c] | 3.6[c] | 3.95[c] |
| cff91    | 2.7    | 0.72       | 3.8    | 4.8    |
| cvff     | 3.1    | 0.98       | 4.0    | 6.7    |

[a] Ref. 40. [b] Ref. 41. [c] Refs. 42, 43.

For mica or other polar solids, the interplay between the attractive Coulomb potential and an appropiate repulsive potential plays the major role. To date, no useful force field is available for mica, since often ill-defined partial charges have been used or even full formal charges.[14-23] The partial charges are important to construct a physical potential energy surface, which in turn yields sensible interfacial properties in the simulation. We solved this problem by assigning physically justified atomic charges based on a novel procedure.[25] The significant Coulomb interactions in such polar solids are one to two orders of magnitude higher than the usual dispersive interactions in organic nonpolar systems. This requires a stronger repulsive energy on the same scale. If we had not to consider the presence of organic residues, an exponential potential like a Morse function or a very "hard" distance dependence of the repulsive energy in excess of $r^{-14}$ would be adequate. However, using the common $r^{-9}$ dependence of cff91 to take advantage of precise hydrocarbon modeling, we are forced towards approximations for the inorganic solid. Both the coefficients $A_{ij}$ of the repulsive energy and the equilibrium bond lengths $r_0$ in the polar mineral must be increased in order to counteract the Coulomb attraction properly (see eq 2 below). If solely the $A_{ij}$ are increased to very high values (~100 times), the layers of the phyllosilicates start artificial translations in horizontal direction, as in the attempts of cff91 extension by Hill and Sauer[16] and Teppen et. al.[17]

Due to the nature of the polar solid, we can make simplifications in the energy expression of cff91. (1) We need only the quadratic bond stretching and quadratic angle bending, cutting all cubic and quartic contributions to zero, due to the overall diminished significance of bond and angle terms to the total potential energy. (2) Torsions and out-of-plane interactions are unphysical in clay minerals and can be cancelled together with all cross-terms, which contribute less than 0.1 % to the total potential energy. The force field terms relevant for mica or other clays are, therefore,

$$E_{pot} = \sum_{ij \text{ bonded}} \frac{1}{2} K_r (r - r_0)^2 + \sum_{ijk \text{ bonded}} \frac{1}{2} K_\theta (\theta - \theta_0)^2 \qquad (2)$$

$$+ \sum_{\substack{ij \text{ nonbonded} \\ (1, 3 \text{ excl})}} \left( \frac{A_{ij}}{r_{ij}^9} - \frac{B_{ij}}{r_{ij}^6} \right) + \frac{1}{4\pi\varepsilon_0\varepsilon_r} \sum_{\substack{ij \text{ nonbonded} \\ (1, 3 \text{ excl})}} \frac{q_i q_j}{r_{ij}} \quad .$$

We use the experimental crystal structure[27, 46] for a first guess of the bond lengths $r_0$ and bond angles $\theta_0$. The definition of several force-field types seems important because the variation of bond lengths and angles between the same connected elements in silicates is considerable.[47] The constants $K_r$ and $K_\theta$ for harmonic bond stretching and angle bending can be derived from the approximate IR frequencies of 1000 and 500 $cm^{-1}$,[48] corresponding to approximately 600 kcal/(mol·Å$^2$) and 80 kcal/(mol·rad$^2$), respectively. The initial Lenard-Jones parameters are taken from the universal force field[49] because these values are consistent with periodic correlations: nonbond equilibrium distances $r_0$ with atom size and dispersive energies $E_0$ with polarizabilities.[50, 51] The atomic charges are taken from ref. 25 where their derivation is thoroughly described.

We carried out a global optimization of the adjustable parameters towards an optimum geometric reproduction of mica, using a box of $4 \times 3 \times 2$ unit cells with a multipole cutoff[52] in a series of molecular-mechanics energy minimization.[35] The resulting force field produces a minimum of acting forces on the equilibrated structure and gives fast convergence. The final parameters are summarized in Table 2.

As a test of the force field, we applied it to our $4 \times 3 \times 2$ mica structure with ~3000 atoms. The fit to the experimental crystal structure[27] is very good: The rms spatial deviation over all atoms amounts to 18 pm in molecular mechanics and to 24 pm in molecular dynamics at ambient temperature. No signs of distortions or artificial movements are present; an equilibration time of ~30 ps is needed to obtain constant energies. As we stated above, no alterations are made to cff91 for the organic residues.

**Table 2:** Extension of cff91 for mica. Parameters for bonds ($r_0$ in Å, $K_r$ in kcal/(mol·Å²)), angles ($\theta_0$ in degrees, $K_\theta$ kcal/(mol·rad²)), van-der-Waals interaction ($r_0$ in Å, $E_0$ kcal/mol), and atomic charges $q_i$. See eq 2.

| | Bonds | Angles | Nonbond | | | | |
|---|---|---|---|---|---|---|---|
| | | | K | Al | Si | O | H |
| $r_0/\theta_0/r_0$ | exptl · 1.07 [a, b] | exptl [a, b] | 4.2 | 4.7 | 4.7 | 3.98 | 2.5 |
| $K_r/K_\theta/E_0$ | 600 | 80 | 0.035 | 0.50 | 0.40 | 0.06 | 0.02 |

| | Tetrahedral layer | | Octahedral layer | | | Interlayer | | |
|---|---|---|---|---|---|---|---|---|
| | Si (Al-defect) | O (when bonded to Al-defect) | Al | O (when bonded to H) | H | $K^+$ | N | C bonded to N |
| Charges $q_i$ | 1.1 (0.8) | −0.55 (−0.78333) | 1.45 | −0.75833 (−0.68333) | 0.20 | 1.0 | −0.10 | +0.275[c] |

[a] Within mica, 34 bonds and 104 angles are defined. [b] Ref. 27. [c] The charge may be partly spread over the adjacent hydrogen atoms.

### 3.2. Setup of the System and Simulations

The size of our 2D-periodic box is limited by the duration of the simulation. We choose the upper half of $4 \times 3 \times 1$ mica unit cells[27, 46] (see Figure 1). This unit accomodates 12 cations (see Figure 2) and is rotated to have ordinary cubic symmetry by changing the angle $\beta$ in the C 2/c symmetric cell from 95.74° to 90°. Then the mica sheet lies in the $xy$ plane and the lattice direction $c$ is identical with the Cartesian $z$ axis. This operation is permitted because we do not change any relative atomic coordinates and do not aim for periodicity in the vertical ($z$-) direction. Si $\rightarrow$ Al substitution on the side opposite to the quaternary ammonium ions is omitted to limit the box size and maintain charge neutrality; the effects on the simulation are negligible.

The substitution pattern of Al against Si on the mineral surface is important because it determines the distribution of negative charge,[25] which directly influences the position of positive counterions.[53] Between nearest Al atoms, Al-O-Al contacts do not occur, Al-O-Si-O-Al contacts are rare, Al-O-(Si-O-)$_2$-Al contacts form the majority, and the incidence of Al-O-(Si-O-)$_3$-Al or longer nearest connections is again rare.[29, 54, 55] For the substitution ratio Al/Si $= 1/3$, the charges are homogeneously dispersed without long-range order of the Al-defects on the surface. For our small box, we make the concession of a regular para-distribution of the Al over the six-membered rings in the tetrahedral sheet, which reflects the main feature of the distribution (see Figure 1a).[29] The small box size and this idealization must be seen in relation to the real macroscopic structures (see Section 4.5).

Our octadecyltrimethylammonium ($C_{18}$) and dioctadecyldimethylammonium ($2C_{18}$) ions were constructed to agree with the crystal structure of the solid $2C_{18}$ bromide monohydrate.[45] The most reasonable starting conformations are extended, as supported by AFM data[5] and surface force measurements.[4]

We assemble the organic cations on one mica surface and carry out the simulation to an equilibrium state at a given temperature. Then we combine two of the equilibrium structures to form a layered structure, which is subjected again to molecular dynamics until equilibration and subsequent sampling of snapshots. The procedure comprises always a few hundred steps of potential energy minimization on the starting structure and then NVT dynamics with initial velocities from the Boltzmann distribution, velocity-Verlet integrator, 1 fs time step, direct velocity scaling with a temperature window of 10 K for temperature control, using the Discover program from MSI.[35] $\varepsilon_r$ is 1.0. For the summation of the Coulomb energy for each atom, spatial segments of sufficient size and electrical neutrality are essential. The cell multipole method by Ding et. al.[52] is a useful tool in our case (third order, 2 layers of cells). The simulation boxes are always at least 2 nm larger than the alkyl-mica system in the vertical direction to facilitate free movement of the chains and adjustment of equilibrium basal-plane spacings in the duplicate structures. Employing this machinery, mica or related minerals with organic structures can be simulated.

## 4. Results of Molecular Dynamics Simulations for $C_{18}$-Mica and $2C_{18}$-Mica

We performed molecular dynamics on single and duplicate mica sheets where 12 and 24 organic cations are attached, respectively. The structures are equivalent to complete alkali-ion exchange, periodic in the $x$ and $y$ directions, and investigated at 20 °C and 100 °C to examine the structure and to surmise on the phase transitions in this range.[12, 13]

At equilibrium, the structures are characterized by thermal fluctuations around a mean value. We employed always several starting structures that converged to the same average structure.

Equilibration was possible after 150 ps in the best cases, but simulations up to 3 ns were usually done. All changes with temperature reported below are fully reversible upon reversal of the temperature to its original setting. Therefore, we assume that they reflect equilibrium changes.

### 4.1. Arrangement of the Ammonium Groups on the Surface

In any ammonium-mica structure, the ammonium groups prefer positions above the surface cavities. The pattern on the surface depends on the short-range ordered Al distribution: positions in cavities surrounded by two Al atoms are preferred over positions in cavities surrounded by only one or zero Al atoms. For both, $C_{18}$-mica and $2C_{18}$-mica, the nitrogen atoms of the ammonium ions sit 380 ($\pm$10) pm (20 °C) to 390 ($\pm$10) pm (100 °C) above the plane of the superficial silicon and aluminum atoms. In the case of $2C_{18}$ ions, the side arm of the second alkyl chain may have different orientations on the surface for each ion.

The positions of the ammonium nitrogens change remarkably upon heating in $C_{18}$-mica. The $C_{18}$ ions are largely confined to their initial location at room temperature, but they move readily across the surface cavities to form new arrangements when the temperature is higher (see Figures 3 and 4). In $2C_{18}$-mica, the ions are strictly confined at both temperatures; the higher average number of alkyl chains per surface area imposes sufficient geometric restraints. Compared to unsolvated alkali ions, both $C_{18}$ ions and $2C_{18}$ ions (when not very close-packed) can move from one cavity to another with less activation energy because the positive charge is spread over the neighboring atoms to nitrogen, lowering the "electrostatic" crossing barrier.

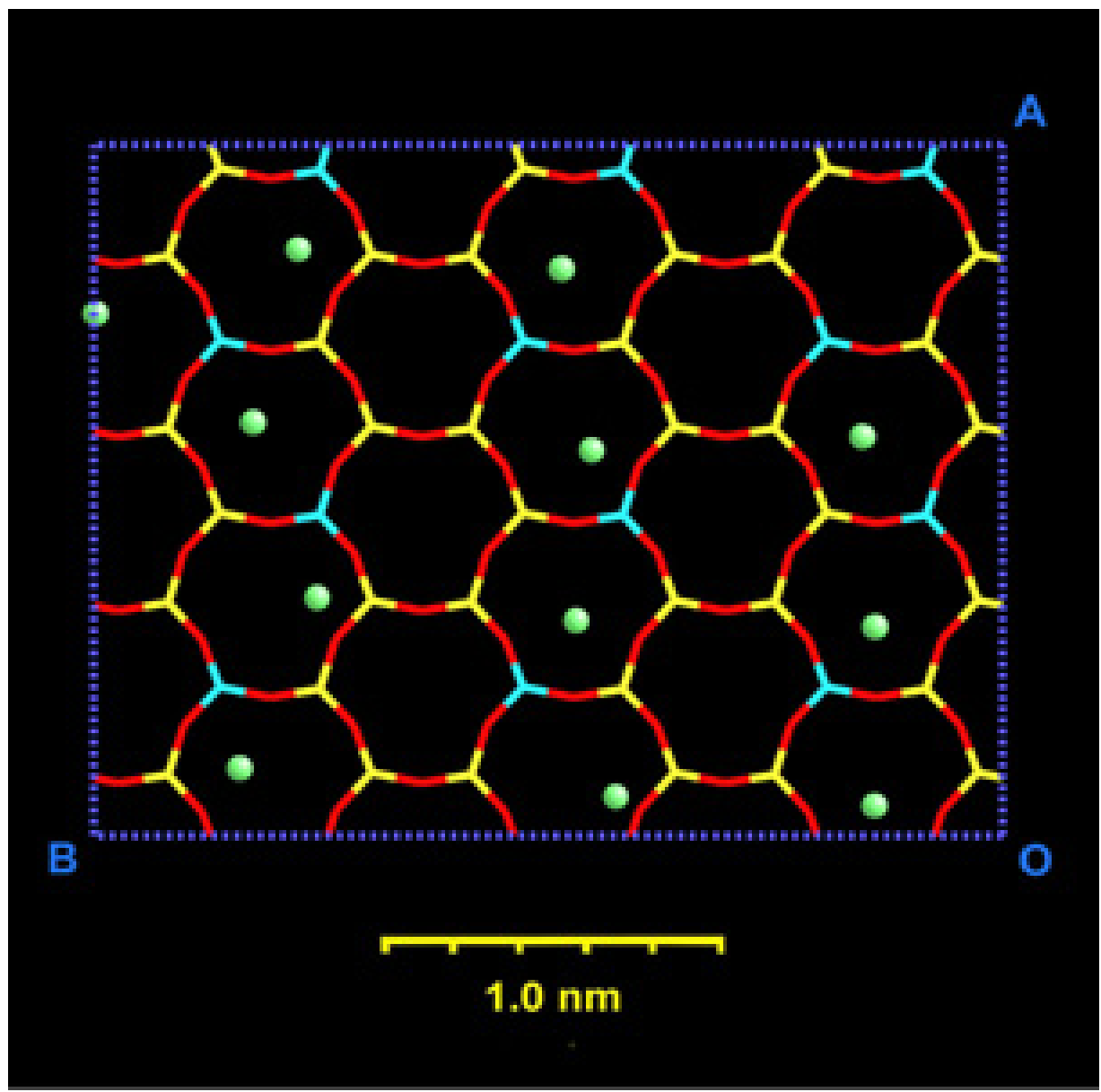

(a)

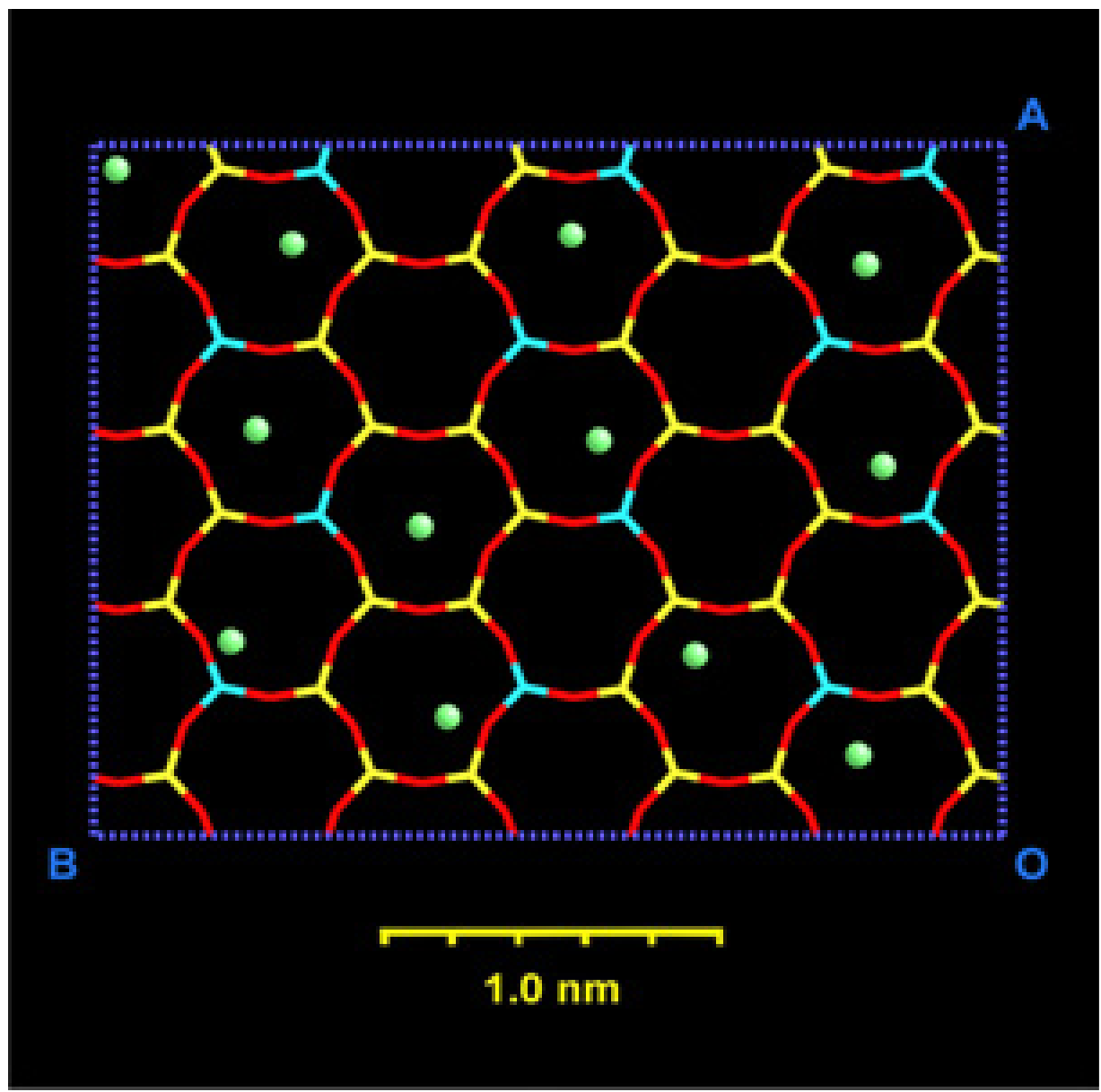

(b)

**Figure 3.** Distribution of $C_{18}$ ions on the mica surface, represented by the green nitrogen atoms, after 400 ps of MD simulation (a) at 20 °C, (b) at 100 °C. With increasing temperature, $C_{18}$ ions can move across cavities. For $2C_{18}$ ions, this is not possible and their arrangement is similar to Figure 1a at both temperatures.

## 4.2. Tilt Angles of the Alkyl Chains

The orientation of the alkyl chains with respect to the *a* or *b* (*x* or *y*) axes does not show a clear preference. It may depend on the Al substitution pattern on the surface and the orientation of the side-arms in case of $2C_{18}$-mica. In contrast, the chains tilt relative to the surface normal with a rather constant angle $\theta$ for a given structure, thus optimizing the van-der-Waals interactions. The tilt angles are essentially the same for single and layered mica sheets. Following Nuzzo et. al.,[56] we define the tilt angle $\theta$ as the angle between a fitted straight line along the carbon atoms of an extended alkyl chain and the projection normal to the surface. $\theta$ is the average over all 12 alkylammonium ions in 10 independent snapshots of the single-layered mica sheets after equilibration. As shown in Table 3, the inclination angle is largest (55°) in $C_{18}$-mica at 20 °C (see Figure 4a). At 100 °C, the system is "molten" (see Figure 4b) and the inclination angle cannot be defined. In $2C_{18}$-mica, we obtain a tilt angle of 30° at 20 °C, which is smaller than in $C_{18}$-mica due to dense packing because of the second alkyl arm (see Figure 4c). At 100 °C, the system still exhibits a rather strict order and the tilt angle is obtained at 13°. The value decreases with increasing temperature because more *gauche*-conformations exist at higher temperature and the effective "thickness" of the chain increases (see Figure 4d).

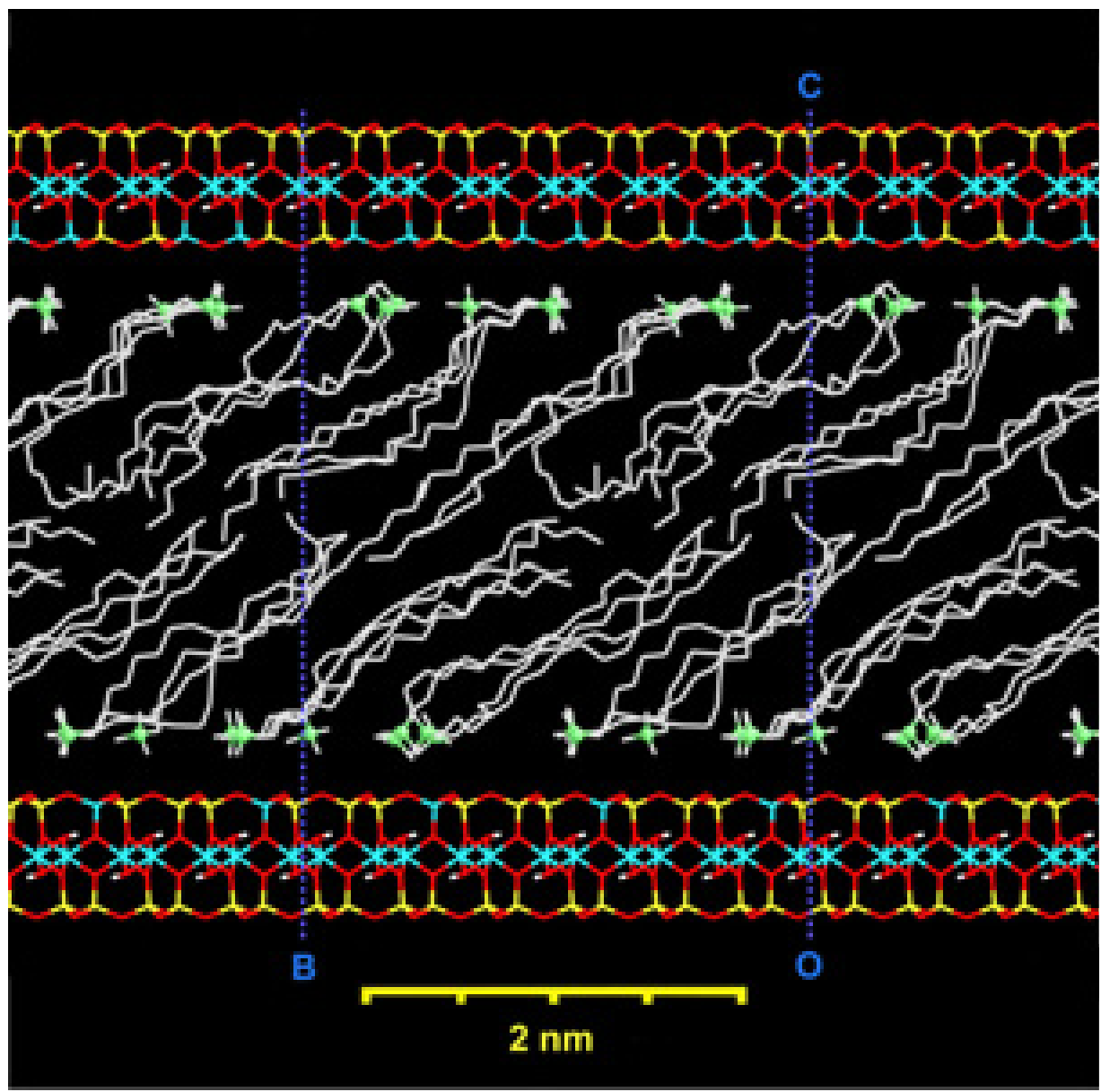

(a)

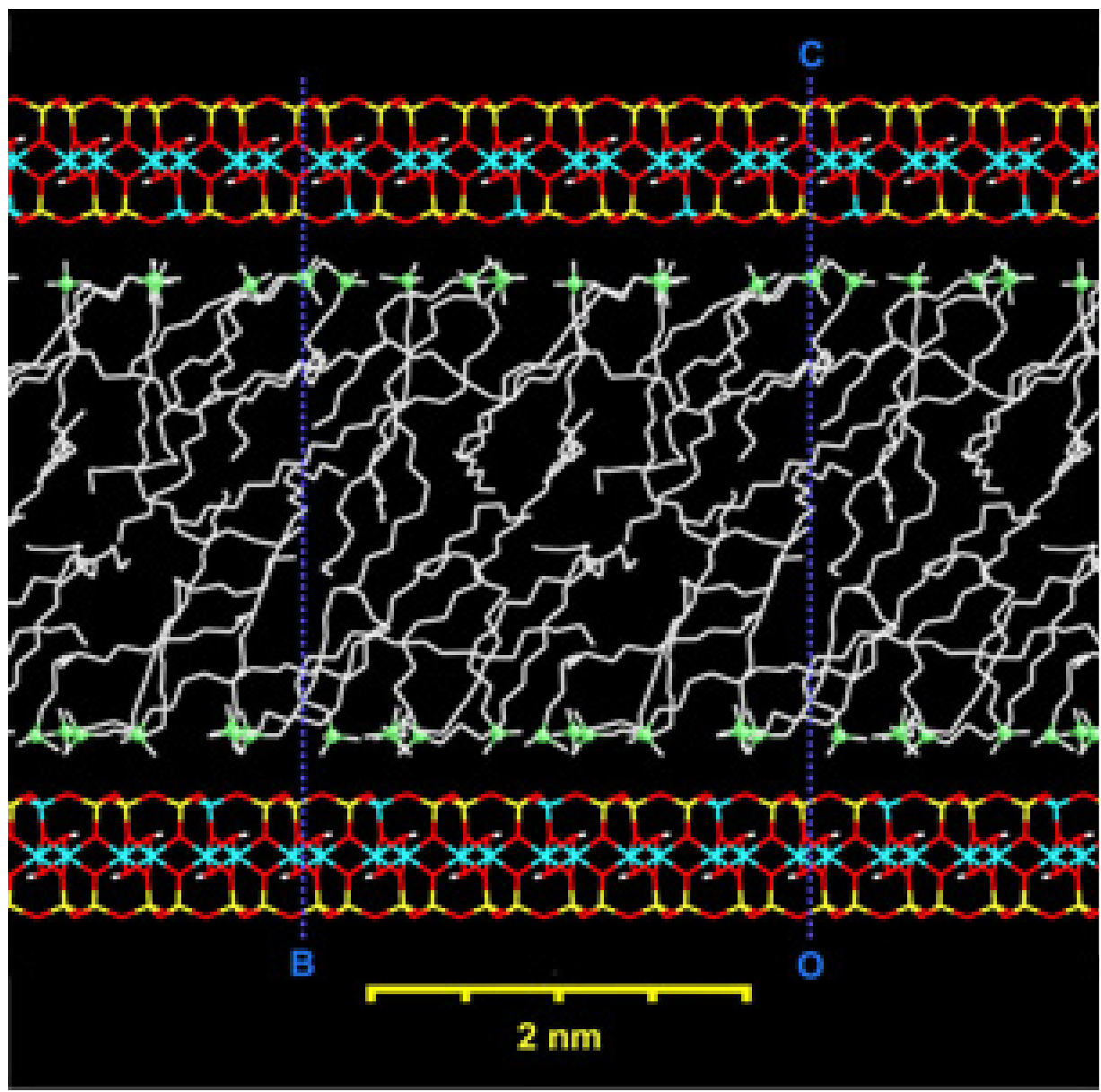

(b)

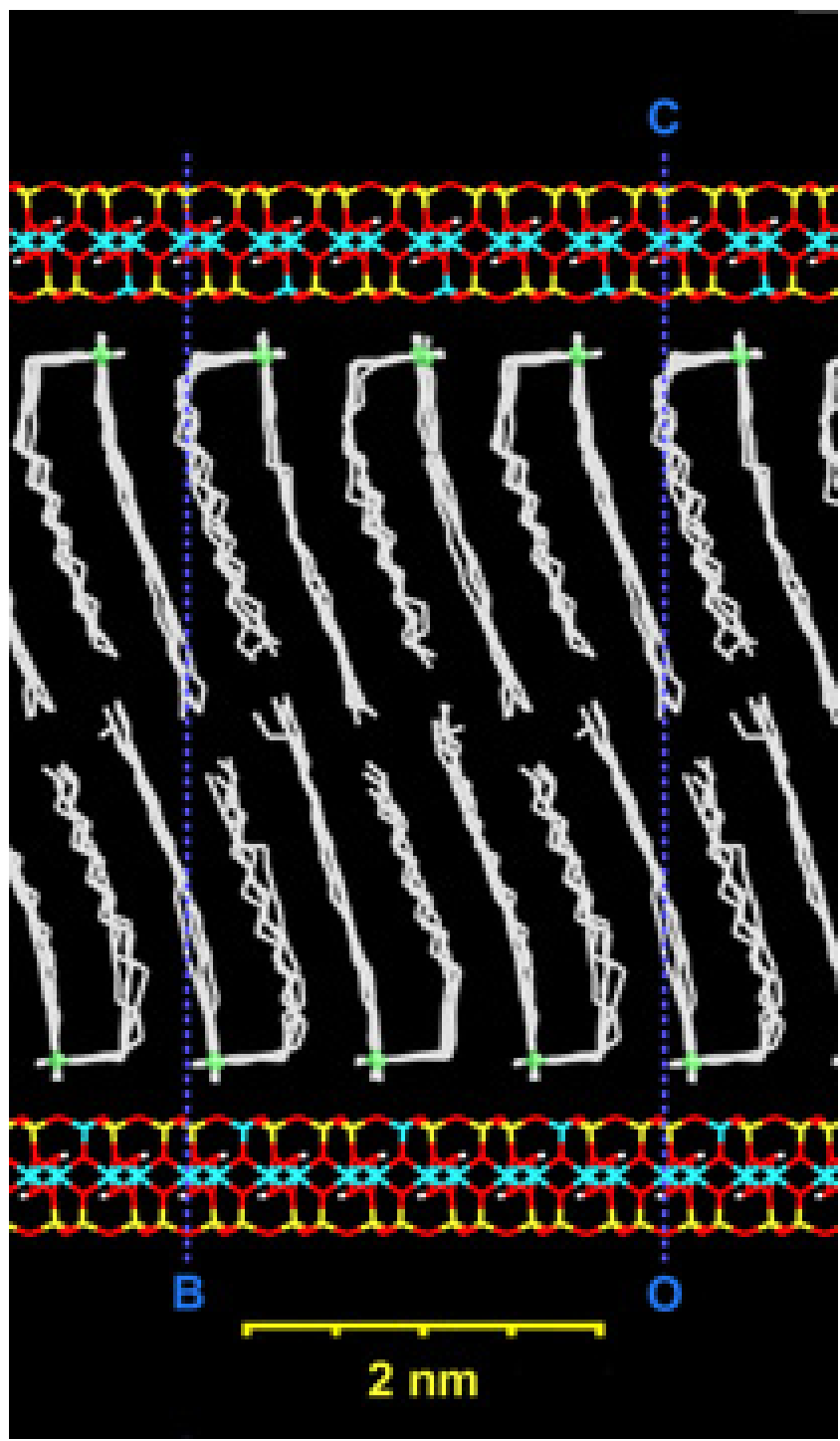

(c)

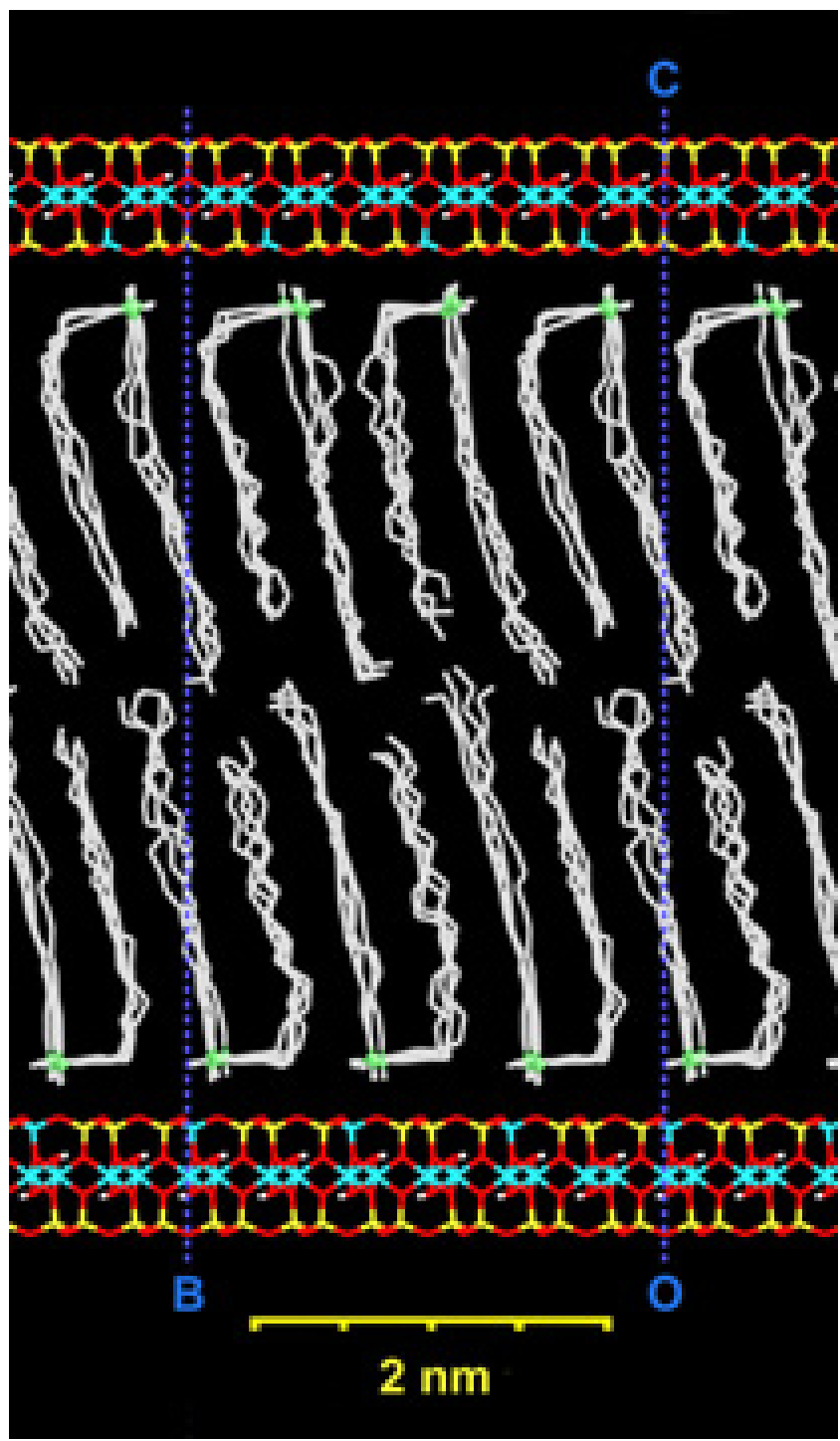

(d)

**Figure 4.** Snapshots of $C_{18}$-mica and $2C_{18}$-mica in MD after 400 ps, viewed along the *a*-direction. (a) $C_{18}$-mica, 20 °C; (b) $C_{18}$-mica, 100 °C; (c) $2C_{18}$-mica, 20 °C; (d) $2C_{18}$-mica, 100 °C. A major conformational change (corresponding to two phase transitions) in $C_{18}$-layered mica can be seen, whereas in $2C_{18}$-mica only some more *gauche*-conformations are present at 100 °C and no order-disorder transition occurs due to close packing.

### 4.3. Basal Plane Spacing and Architecture of the Duplicate Structures

The basal plane spacing can be deduced for the layered structures (see Figure 4). It is given as the difference between the average $z$ coordinates of the octahedrally coordinated Al atoms (see Figure 1) in the two mica sheets, and constitutes a measurable quantity (e. g., by XRD). We state the average over 100 snapshots at a time interval of 1 ps, after the system's energies are equilibrated. For $C_{18}$-mica with quantitative cation exchange, we obtain a slight increase in basal plane spacing from 3.67 nm to 3.82 nm when the temperature is higher. For $2C_{18}$-mica with quantitative cation exchange, the basal-plane spacing of 5.40 nm increases to 5.63 nm upon heating (see Table 3).

The relative orientation of the alkyl chains (given by the abovementioned fitted straight line through the carbons) in both halfs of the layered "sandwich" can be different. In the presented structures (see Figure 4), the alkyl chains arrange with a common director. They may also form a "fishbone" pattern without a significantly higher internal energy ($\pm$ 4 kJ per mole $2C_{18}$). The basal-plane spacing is then about 0.05 nm to 0.10 nm higher. A wide distribution of orientations (from 0° to 360°) between the two parallel mica platelets seems possible.

 

**Table 3:** Results of the molecular dynamics simulations and comparison with experiments. Tilt angles are given in degrees and basal plane spacings in nm.

| | | Single structures | | Layered structures | | |
|---|---|---|---|---|---|---|
| | | Tilt angle to the surface normal | | Basal plane spacing in sand-wiched structures | | Number of *gauche*-torsions per ion |
| | | MD | exptl[a] | MD, 100 % substituted | XRD, 70-80 % substituted | (MD) |
| $C_{18}$-mica | 20 °C | 55 ±2 | 55 ±3 | 3.67 ±0.02 | 3.89 ±0.02[b] | 3.8 ±0.1 |
| | 100 °C | - | - | 3.82 ±0.02 | 4.02 ±0.02[b] | 4.9 ±0.1 |
| $2C_{18}$-mica | 20 °C | 30 ±2 | 30[c], 38 | 5.40 ±0.03 | 4.80 - 4.70[d] | 5.3 ±0.1[e] |
| | 100 °C | 13 ±3 | - | 5.63 ±0.03 | 5.04 ±0.02[d] | 7.2 ±0.1[e] |

[a] Ref. 9. [b] Ref. 12. [c] See section 5.2. [d] Ref. 13. [e] There are 4 intrinsic *gauche*-arrangements at the junction of the two alkyl chains via the nitrogen, included in these numbers.

### 4.4. *Gauche*-Arrangements

The conformations of the chains can be conveniently analyzed. As shown in Figure 4 a/b, the $C_{18}$ chains are "melting" when the temperature is increased from 20 °C to 100 °C, in combination with the above mentioned rearrangements on the surface. The $2C_{18}$ chains retain their order with increasing temperature, although the irregularities in the chain conformation are more frequent and the basal-plane spacing increases (see Figure 4 c/d).

These effects can be quantitatively described by the number of *gauche*-torsions ($C - C - C - C$ torsion angle in the range $[-120°, +120°]$) per alkylammonium ion. We count them along the backbone from the terminal C atom towards the N atom and going further along the chain towards the second terminal C atom in the case of $2C_{18}$ ions. We calculate average values over all chains in 100 snapshots after equilibration. In $C_{18}$ ions at 20 °C, 3.8 out of 16 torsion angles are *gauche*, and at 100 °C the number of *gauche*-conformations is increased to 4.9 (see Table 3). In $2C_{18}$ ions at 20 °C, 5.3 out of 34 torsion angles are *gauche*, whereby four of them arise from the presence of the side-arm of the second alkyl chain and are already present in the solid $2C_{18}$ bromide structure.[45] At 100 °C, the number of *gauche*-torsions increases to 7.2. If we consider only the extended parts of the $2C_{18}$-chains, the number of *gauche*-torsions increases from 1.3 to 3.2 upon heating. Thus even at 100 °C, there are only 1.6 *gauche*-arrangements per $C_{18}$ chain leading to disorder in $2C_{18}$ ions, which is less than half the 3.8 *gauche*-arrangements in $C_{18}$ ions at 20 °C. With this small fraction of *gauche*-arrangements in $2C_{18}$ ions, an order-disorder transition is unlikely.

It should be noted that these values depend critically on the surface charge. For a $Si_{3.06}Al_{0.94}$ composition of the tetrahedral layer instead of $Si_3Al_1$,[5] e. g., the available surface area on mica per alkyl chain is by 6 % greater and may increase the number of *gauche*-incidences noticeably.

### 4.5. Reliability of the Data

The extended cff91 has a high precision and geometric deviations in the reproduction of the mineral with attached alkyl chains are in the order of 1 to 5 percent (rms deviation in scaled coordinates). While the reproduction of the mineral is precise within 1 % in scaled coordinates, we assume that the cff91 could be improved for the reproduction of the organic part, eventually employing an $r^{-4}$ distance dependence for the dispersive van-der-Waals energy.[38]

Moreover, we assume a regular Al substitution pattern. This leads to a higher order of the superficial chains and makes processes such as the rearrangements of the $C_{18}$ ions more clearly visible. However, we find a tendency towards ideal close packing then, e. g., an optimized interdigitation of the chain ends in the double structures, and should be aware that real, μm-sized mica sheets carry more than one million tethered ions and their minimal closest approach determines the basal plane spacing.

In conclusion, the main sources for deviations are the force field and the higher degree of order in the small periodic box compared to a real system. There is a systematic trend to underrate the basal-plane spacings by few percent (see Section 5.1), but within this small deviation all data are reliable.

# 5. Juxtaposition with Experiment[10-13]

In this section, we want to combine the available experimental data[4-7, 9-13] and our simulation results for structural understanding of the modified mica, including less than 100 % ion exchange. We explain the previously reported phase transitions[12, 13] and order the results by geometry considerations. We also explain the mechanism of ion exchange as far as possible and propose a thermal elimination mechanism.

## 5.1. $C_{18}$ on Mica: Structure

From the work of Hayes and Schwartz,[10] as well as Fujii et. al.,[11] we know that $C_{18}$-ions form islands on the mica surface, i. e., the alkali ions and the organic ions are separated in different phases, even if the overall degree of ion exchange is 85 % or higher. The areas with $C_{18}$ islands are 100 % ion-exchanged and the zones between them contain almost exclusively alkali ions. Therefore, our simulated $C_{18}$-mica corresponds to the inner part of such island structures. The simulated tilt angle of 55° (see Table 3) matches exactly the result from NEXAFS spectroscopy,[9] for which 55 ° is a special case where no scaling parameter is required.

The basal-plane spacing in the experimental XRD patterns[12] is indifferent to the degree of cation exchange (see Table 4); the basal-plane spacing remains at a constant value of 3.9 nm. In the simulation, which is equivalent to 100 % cation exchange, we obtain 3.7 nm, which is within the combined limits of accuracy (see Section 4.5). This finding is consistent with the occurence of islands.[10] The basal plane spacing is solely defined by the height of the separate $C_{18}$ phases on the surface (see Figure 5). The regions containing the small alkali ions (alkali not shown in Figure 5) are "empty spaces" that decrease the density of the conglomerated material.

**Table 4:** Basal-plane spacings (in nm) at ambient temperature for different degrees of alkali exchange. In $C_{18}$-mica, no dependence on ion exchange is visible. In $2C_{18}$-mica, the basal-plane spacing increases roughly linear.

|                | 50 %        | 70-80 %        | 100 % (MD) |
| -------------- | ----------- | -------------- | ---------- |
| $C_{18}$-mica  | 3.9[a]      | 3.9[a]         | 3.7        |
| $2C_{18}$-mica | 3.9[b]      | 4.7-4.8[c]     | 5.4        |

[a] Ref. 12. [b] Estimated (see text). [c] Ref. 13.

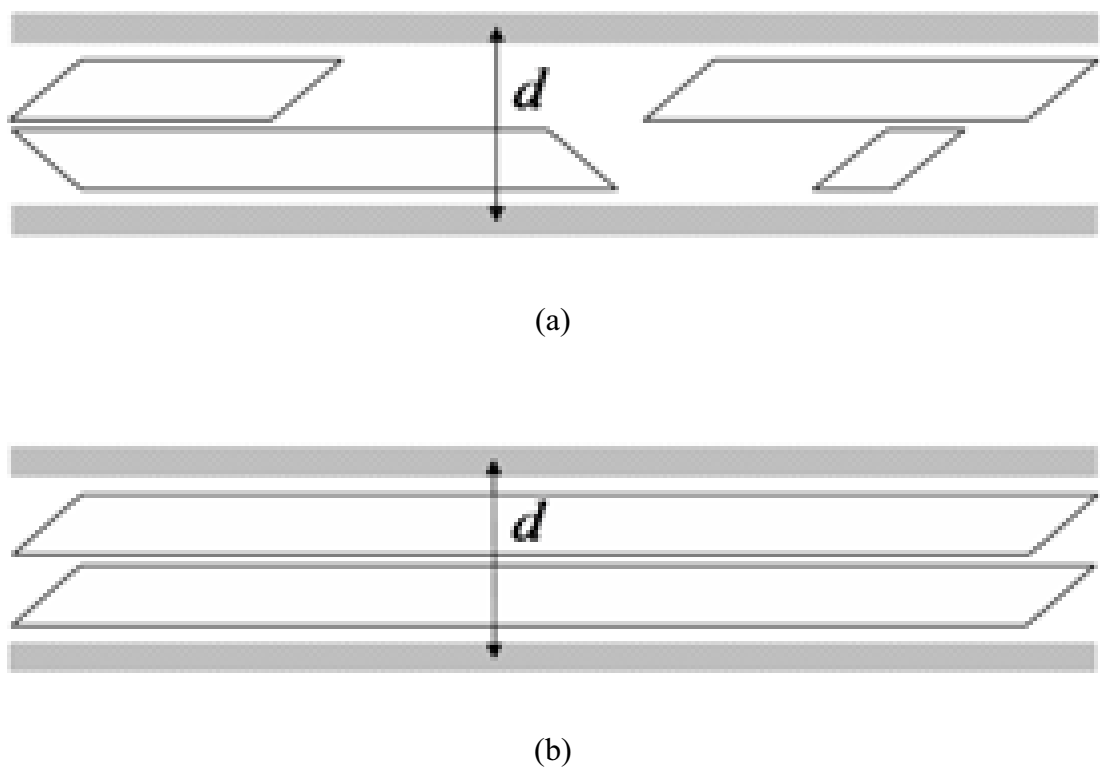

(a)

(b)

**Figure 5.** Sketch of $C_{18}$-mica conglomerates. Phase separation in case of nonquantitative alkali exchange (a) leads to the same basal plane spacing $d$ as in case of quantitative alkali exchange (b). The inclination direction of the islands may vary.

From the computed value of 3.8 *gauche*-arrangements per $C_{18}$ chain in the islands and a representative snapshot of the system (see Figure 4a), we recognize a considerable deviation from an all-*trans* conformation at 20 °C. It is, therefore, not surprising that there is a phase transition commencing at 40 °C in experiment[12] and in the simulation.

### 5.2. 2$C_{18}$ on Mica: Structure

For quantitative ion exchange, we found inclination angles of the chains around 30°. A tilt angle of 38° was reported from NEXAFS spectroscopy[9] with relative scaling to a hexadecane thiol-on-gold standard, for which 33° was assumed. Regarding this reference system, a range of val-ues between 20° to 40° has been reported in the literature.[56, 57] We tend to trust the suggestion of 25° by Porter et. al.,[57] which is based on measurements on a whole series of homologous alkane thiols with different techniques. This implies an inclination angle of ~30° for $2C_{18}$-mica, in agreement with our simulation. In contrast to $2C_{18}$-mica, an absolute assignment of the tilt angle by NEXAFS was possible for $C_{18}$-mica,[9] which matches our computed value (see Section 5.1).

For less than 100 % ion-exchanged $2C_{18}$-mica, no structural propositions have been made, although some measurements are available.[13] The computed basal-plane spacing of 5.40 nm in 100 % alkali-exchanged $2C_{18}$-mica is clearly above the value of 4.7 to 4.8 nm in 70-80 % cation-exchanged $2C_{18}$-mica (see Tables 3 and 4). Assuming a homogeneous coverage of all surface area, we can guess an expectation of 3.9 nm for 50 % cation-exchanged $2C_{18}$-mica because it has the same number of alkyl chains as found in 100 % ion-exchanged $C_{18}$-mica per surface area, while the effects of small alkali ions and ammonium head groups may be neglected. A linear relationship between the degree of alkali exchange and the basal plane spacing is then obeyed (see Table 4), which strongly suggests a homogeneous mixture of alkali ions and $2C_{18}$ ions. We provide some evidence for this conclusion: (1) The density of the hydrocarbon moiety is a constant[50] and the effect of the small alkali ions is negligible; if then additional exchange of alkali against $2C_{18}$ rises the basal-plane spacing, this additional $2C_{18}$ could not be built into an empty space (refer to Figure 5) but rather into an existing homogeneous layer of less than 100 % exchange, which adjusts to the incoming mass by augmentation of its volume. (2) The enlarged XRD peaks for 70-80 % ion-exchanged $2C_{18}$-mica seem to be a superposition of more than one definite structure[13] and indicate possible deviations from ideal homogeneity in the single-phase layer. (3) The density of alkyl chains per surface area is double as high in $2C_{18}$-mica with 20 to 30 % alkali ions compared to $C_{18}$-mica with 20 to 30 % alkali ions. This may foster a uniform phase because it gives more conformational freedom to the alkyl chains. (4) Moreover, we find only one reversible phase transition upon heating.[13] The alkyl chains occupy double as much space on the mica surface per head group compared to $C_{18}$-mica, so that rearrangements of the head groups and the resulting metastability after fast cooling are minimal (see Section 5.3).

These arguments suggest that the alkali ions are "randomly" interspersed between the $2C_{18}$-ions in $2C_{18}$-mica with 70 to 80 % ion exchange. (A degree of alkali exchange less than ~50 % might, however, favor island formation.)

### 5.3. $C_{18}$ on Mica: Phase Transitions

In $C_{18}$-mica with complete alkali exchange, or a separate $C_{18}$ phase on mica, the simulation indicates a major conformational change due to "melting" of the templated chains (see Figure 4 a/b) and an increased mobility of the $C_{18}$-ions across the surface cavities (see Figure 3) at elevated temperature. With the limited temperature precision in the simulation, we cannot decide if the entire changes observed in the simulation are brought about by one or more successive phase transitions. From the experimental observations (see below in this section),[12] however, we conjecture that there are two separate phase transitions: one transition is due to the conversion of the disordered $C_{18}$ rods into broken $C_{18}$ rods, and the second transition is due to rearrangements of the ammonium head groups on the surface with a concomitant conversion of the broken $C_{18}$ rods into a coil-like structure. Altogether, a high degree of disorder is introduced. However, the chains remain essentially at the same locations once they have equilibrated and the number of *gauche*-conformations does not increase significantly (Table 3). The first phase transition seems to occur when the number of *gauche*-incidences per chain reaches a threshold value around 4.0 out of 16 torsional angles. The enthalpy of a transition, $\Delta H_m = \Delta S_m \cdot T_m$, is then associated with the entropy increase $\Delta S_m$ of the system by breaking down the regular order of the chains during "melting" while the number of *gauche*-arrangements remains approximately steady.

Experimentally, in $C_{18}$-mica with 50 to 80 % alkali ion exchange, two distinct phase transitions have been observed on heating and metastable phases were obtained on fast cooling after the second transition.[12] Since the $C_{18}$ ions are always arranged in islands,[10, 11] the situation is analogous to 100 % ion exchange. Therefore, we surmise that the first phase transition at 40 °C is due to breaking of the disordered $C_{18}$ rods in the islands (see Figure 4a), and the second transition at 60 °C is due to rearrangements of the $C_{18}$ ions on the surface, concertedly leading to more conformational freedom of the whole backbone and a coil-like disordered structure. In addition to these findings, there is evidence that the phase boundaries between islands and alkali ions are preserved. On fast cooling, the structures resulting from the second transition at 60 °C "freeze" into a metastable disordered-rod structure, where the head groups have not returned to their original positions. This metastable system exhibits one immediately reversible phase transition around 40 °C similar to the first recorded transition until the reverse rearrangements to yield the original structure have taken place (after several hours).

We provide detailed evidence in the following. Firstly, we reason that all transitions occur within the islands and that the shape of the islands is essentially preserved. Since pure mica does not exhibit phase transitions until very high temperatures are reached and $C_{18}$ ions forms separate phases on mica,[10, 11] we assume that the phase transitions of $C_{18}$-mica originate in the $C_{18}$ islands. We mentioned that alkali ions have much less mobility on the surface than the alkylammonium ions (see Section 4.1) so that they cannot leave the cavities without the help of a polar solvent. As a consequence, also the alkylammonium ions are fixed in proximity to their initial locations. Otherwise, excess negative charges would be created where the alkylammonium ions leave and excess positive charges where they migrate to. Rearrangements are, therefore, limited to the neighbor cavities. The basal-plane spacing increases slightly on heating in the experiments[12] (see Table 3), in accordance with preservation of an island structure. If the separate phases were coalesced after the transition, we would expect a decrease of the basal-plane spacing (see Section 5.2). According to these arguments, the basic shape of the islands is preserved and no significant extension of the boundaries occurs during the transitions.

Secondly, we discuss the phase transitions. For the first event at 40 °C, the data from Osman et. al.[12] suggest a transition enthalpy of less than 6 kJ per mol $C_{18}$, which is ~12 % of the melting enthalpy of $n$-$C_{19}H_{40}$ at 32 °C.[50] The comparatively small corresponding entropy gain (and the slightly elevated transition temperature in $C_{18}$-mica compared to $n$-$C_{19}H_{40}$) can be related to the tethering of the chains. The changes are also not sufficient to be detected in the IR spectrum and the density is almost constant (no increase in basal-plane spacing).[12] Therefore, we conjecture that the transition corresponds to a breaking of the disordered rods, i. e., a partial melting of the chain backbones, while rearrangements of the head groups are less likely at only 40 °C.

The transition enthalpy at 60 °C is 12 kJ per mole $C_{18}$, compared to only 6 kJ/mol for the transition at 40 °C,[12] which means that the entropy gain due to both transitions amounts to roughly one third of the melting entropy of $n$-$C_{19}H_{40}$. Also, a change in the IR spectrum occurs as in the melting of nonadecane, although to lesser extent,[12] (the $C-H$ stretching vibrations in the $-CH_2-$ groups shift to slightly higher energies) and the density decreases (increase of the basal-plane spacing). When the heated samples with 50 to 80 % ion exchange are cooled at 10 °C/min, several hours at room temperature are required to recover the transition at 60 °C again; the enthalpy of the second transition increases with standing time.[12] These experimental arguments require a substantial increase of disorder of the $C_{18}$ chains: when the temperature increases above 40 °C, the broken $C_{18}$ rods increase their internal energy further until a new threshold for expansion is reached. As we have seen in the simulation (see Figure 3), rearrangements of the head groups occur and at the same time the chain backbones obtain more conformational freedom to adopt coil-like conformations (see Figure 4b). We believe this constitutes the transition at 60 °C and accounts for the higher melting enthalpy, the IR spectrum, as well as the density increase. We obtained indications in the simulation that rearrangements are slow at ambient temperature (Figure 3a). Therefore, we can understand the generation of a supercooled metastable phase after fast cooling; the chains reorganize themselves slowly by reverse rearrangements of the ammonium headgroups.

Finally, the metastable form of $C_{18}$-mica with ~80 % ion exchange, which is obtained on fast cooling of the product of the second transition at 60 °C, undergoes one reversible melting transition near 40 °C on heating and cooling with approximately ±7 kJ per mol $C_{18}$.[12] This transition is similar to the first recorded transition at 40 °C, but the DSC peak is slightly broader and $\Delta H_m$ is slightly higher.[12] The transition is likely to proceed between a disordered-rod structure and the coil-like structure. The difference to the first recorded transition is, however, that the ammonium head groups are still in a favourable position for the partially molten (coil-like) structure but not for the disordered-rod structure. This explains the less-well defined DSC peak as well as a tendentiously higher melting enthalpy and melting entropy. The reverse rearrangements of the ammonium headgroups may even be incomplete after several hours.[12]

**5.4. $2C_{18}$ on Mica: Phase Transitions**

In $2C_{18}$-mica with 100 % alkali exchange, no phase transition can be discerned on heating in our simulation (see Figure 4 c/d). Although the ordering of the chains decreases (Table 3), the number of *gauche*-incidences in the extended parts of the $C_{18}$ chains is not enough to trigger an order-disorder transition. We estimated that 4.0 torsional angles should be *gauche* in the extended part of a $C_{18}$ chain to facilitate a phase transition, whereas in the geometrically restrained $2C_{18}$-mica only 0.8-1.6 angles are *gauche* in a $C_{18}$ backbone between 20 and 100 °C (see Section 4.4).

In $2C_{18}$-mica with 70 to 80 % alkali-ion exchange[13] or with decreased Al/Si substitution,[5] one reversible phase transition occurs at ~53 °C.[5, 13] It is due to a conversion of disordered rods into tethered coils without significant head group rearrangements on the surface.

The evidence is as follows. As we derived in Section 5.2, the $2C_{18}$ ions are mixed with interspersed alkali ions in certain domains. The melting transition proceeds smoothly and quickly in both directions.[13] The C – H stretching vibrations as well as the $^{13}$C-NMR chemical shifts for the inner methylene groups of the octadecyl chains are changing during the transition.[13] The occurence of a single reversible "melting" transition implies that rearrangements of $2C_{18}$-ions across the cavities rarely occur. The density of alkyl chains on the $2C_{18}$-mica surface is roughly double as high as in $C_{18}$-mica so that the vast majority of the cavities on the mica surface is covered by either alkyl chains or potassium ions. Thus rearrangements, in particular, a second phase transition, are hardly possible in $2C_{18}$-mica. The number of *gauche*-arrangements is probably ~12 per $2C_{18}$ ion at 53 °C, corresponding to the required threshold value of ~4 *gauche*-arrangements in the extended part of each $C_{18}$ backbone (see Section 5.3). The transition is likely to be a "melting" process of the tethered disordered $C_{18}$ rods to a structure between tethered broken $C_{18}$ rods and tethered $C_{18}$ "coils".

## 5.5. Prediction of Structures from Geometrical Arguments

We notice that the occuring structural patterns on mica depend on the number of alkyl chains per unit surface area. With increasing number of alkyl chains per unit surface area, the structural patterns change from phase-separated via intermediate to continuous layers. There is also a correlation with inclination angles in continuous layers (see Table 3).

This aspect relies exclusively on the geometric conditions. We define the surface saturation $\lambda$ as the quotient of the cross-section $A_C$ perpendicular to the outstretched hydrocarbon chains and the surface area $A_S$ available to them:

$$\lambda = \frac{A_C}{A_S}. \tag{3}$$

The cross-section of an all-*trans* hydrocarbon chain perpendicular to its axis is $A_{C, \text{trans}} = 0.188 \text{ nm}^2$ at ambient conditions ($0.175 \text{ nm}^2$ at 90 K),[44, 45] and the area of two cavities on the mica surface is $A_S = 0.468 \text{ nm}^2$ (see Figure 1).[26-28, 30] With an $Al_1Si_3$ composition on the surface, $A_S$ bears one negative charge. $\lambda$ gives an idea what percentage of the surface area would be needed to force the chains to be normal to the surface. We obtain $\lambda_{\text{OTA-mica}} \approx 0.4$, and $\lambda_{\text{DODA-mica}} \approx 0.8$, neglecting the *gauche*-conformations. The value for $C_{18}$-mica is rather low, and it seems not surprising, that islands are preferred structures when the surface saturation is only ~0.3 for 75 % ion exchange. In $2C_{18}$-mica, most of the available surface area is used and a continuous layer is preferred. The value of $\lambda$ indicates which structure will be found on a flat surface (Table 5), regardless of its chemical composition.

**Table 5:** Relation between surface saturation $\lambda$ and surface structures occuring for octadecyl chains.

| $\lambda$ | < 0.4 | 0.4 - 0.6 | > 0.6 |
|---|---|---|---|
| Preferred structure | Islands | Intermediate | Continuous Layer |

Moreover, there is a simple relation between $\lambda$ and the tilt angle $\theta$ in homogeneous layers. When $l$ is the length of the chain, the volume of the chains in the whole layer with (vertically) extended chains is $V_C = A_C \cdot l$. When the chains tilt, their volume remains constant, but their cross-section parallel to the mica surface increases to match the value of $A_S$. According to Cavalieri's theorem, the volume of the chains is now given by the new cross-section and the height $h$ of the chains normal to the surface, $V_C = A_S \cdot h$. From the volume identity and $\frac{h}{l} = \cos\theta$, we obtain:

$$\lambda = \frac{A_C}{A_S} = \cos\theta. \tag{4}$$

With this relation, we predict the tilt angle for $C_{18}$-mica as 66° and for $2C_{18}$-mica as 37° for all-*trans* configured chains. These values are upper limits. If we add $0.024 \pm 0.001$ nm$^2$ to $A_{C,\,trans}$ per *gauche*-torsion in the extended part of the $C_{18}$ chains, we obtain a good fit to all determined tilt angles, with less than 2° deviation (see Table 6). Further effects caused by the bulky quaternary amino group, irregular distribution of the ammonium groups over the cavities, as well as interspersed cations may lead to a real tilt angle a few degrees lower than the calculated value, although these influences become negligible the longer the alkyl chains are.

**Table 6:** Prediction of tilt angles θ (in degrees) of octadecyl chains in continuous layers with the surface saturation λ.

|  | $C_{18}$-mica, 20 °C | $2C_{18}$-mica, 20 °C | $2C_{18}$-mica, 100 °C |
|---|---|---|---|
| λ | 0.597 | 0.870 | 0.968 |
| $\theta = arc \cos \lambda$ | 53.4 | 29.5 | 14.6 |
| Exptl/MD[a] | 55 | 30 | 13 |

[a] See Section 5.1/5.2.

## 5.6. Exchange Mechanism

We want to summarize in short the likely mechanism of formation for alkylammonium mica as far as progress in structural understanding has been made because it is crucial to the understanding of the self-assembly structures. Upon dry cleavage of mica, the surfaces usually have no electroneutrality[58] and the potassium will be scrambled over roughly half the cavities without a high degree of order. When this mica is brought into aqueous solution, we can assume that the distribution of alkali ions, which are $Li^+$ instead of $K^+$ after delamination,[9, 12, 13] changes towards its minimum surface energy (for example, see Figure 1a). Rearrangements of solvated alkali cations will be possible because the positive charge is spread as in tetraalkylammonium ions (see section 4.1.)

When alkali ions on small mica particles are exchanged against alkylammonium ions below the critical micelle concentration (cmc) of roughly $1 \times 10^{-3} M$, the following steps are likely: (1) Firstly monomeric $C_{18}$ or $2C_{18}$ ions are randomly adsorbed onto single mica surfaces. (2) The ions rearrange on the surface to form more stable structures. In the case of $C_{18}$-mica, these are homogeneous islands.[10, 11] If the average number of alkyl chains per surface area rises above a certain value, a relatively homogeneous layer is preferred, as for nonquantitatively exchanged $2C_{18}$-mica with interspersed alkali ions. (3) With elapse of time the single-layered surfaces pair themselves with others to form sandwich-like structures.[12, 13]

The speed of steps 1 and 2 is governed by the temperature and the nature of the exchanged cations (see section 2.2). If $Li^+$ is present, ion exchange takes only hours, whereas with $K^+$ it is not completed after many days at ambient conditions.[10, 11] Termination of step 3 depends on the polarity of the solvent and its removal.

For concentrations of the alkylammonium ions above the cmc, first cylindrical structures and at higher alkyl density homogeneous bilayers are formed.[4, 6] The details about these mechanisms are not yet clear.

## 5.7. A Possible Mechanism for Thermal Decomposition

The thermal decomposition of tetraalkylammonium micas may proceed analogous to the cleavage of quaternary ammonium hydroxides (Hofmann elimination), since the mica-multianion is a medium strong base with the corresponding acid H-mica. The favoured *anti*-periplanar transition state[59] for the β-elimination is not very difficult to achieve. Especially at higher temperature, where the conformational flexibility is high, the first two carbon atoms departing from the nitrogen in an alkyl chain arrange often roughly parallel to the surface, so that conformations promoting the transition state for an E2 elimination are found in the simulation. The β-hydrogens on the ammonium ions approach the level of the upper oxygen atoms of the mica surface about 300 pm, so that H-abstraction may occur. The usual elimination temperature for quaternary hydroxides is 100 to 200 °C.[59, 60] Mica is not as strongly acting as hydroxide ions and, therefore, the elimination temperature is roughly 300 °C.[12, 13] Further evidence could be obtained by a detailed analysis of the elimination products.

## 6. Conclusions

We developed an accurate extension of cff91 for mica, which can be employed for other clay minerals or zeolithes. Although the repulsive nonbond potential should be stronger for highly polar solids than in the given 9-6 formulation, the structures of both the mineral and the organic residues can be adequately modeled.

The computed properties of $2C_{18}$-mica and $C_{18}$-mica are in satisfying agreement with experiments. The ammonium ions occupy positions above the cavities in the mica surface, preferentially with a higher number of Al-defects. The nitrogen atoms reside ~385 pm above the plane of the superficial Si and Al atoms. For less than quantitative alkali-ion exchange, $C_{18}$ on mica forms phase-separated structures, while $2C_{18}$ on mica forms a homogeneous phase mixed with remaining alkali ions. For solid liquid-like phase transitions in a bulk of tethered $C_{18}$ chains, approximately 4.0 *gauche*-arrangements along the backbone are required. The two phase transitions upon heating in partially or fully cation-exchanged $C_{18}$-mica are likely due to breaking disordered $C_{18}$ rods at 40 °C, followed by a transition into a coil-like state at 60 °C with rearrangements of the ammonium headgroups on the surface. On fast cooling, a metastable system is obtained. Reversible "melting" transitions between a disordered-rod state and the coil-like state at ~40 °C are observed until the slow reverse rearrangements to the original structure are achieved. $2C_{18}$-mica with 70 to 80 % ion exchange undergoes only one phase transition at ~53 °C, which is a partial melting of the tethered alkyl chains without significant rearrangements of $2C_{18}$-ions. For 100 % ion exchange, no transition can be discerned in the simulation.

The occurence of island, intermediate, or homogeneously mixed surface structures and the value of the tilt angle in continuous layers are associated with the surface saturation ratio $\lambda$ (see Tables 5 and 6). $\lambda$ allows to make predictions for other surfaces. We suggest that the thermal decomposition of the tetraalkylammonium micas is a Hofmann-type elimination with mica acting as a rigid base.

## Acknowledgements

We express our thanks to Epameinondas Leontidis, Department of Chemistry, University of Cyprus, Walter Caseri, Department of Materials, ETH Zürich, and Andrey Milchev, Department of Physics and Astronomy, University of Georgia, US, for helpful discussions, as well as to Marc Petitmermet, Department of Materials, ETH Zürich, for his continuous support with computational resources. The Swiss Center for Scientific Computing, Manno, Switzerland, also alloted computational resources. Furthermore, the support from the Swiss National Science Foundation (Schweizerischer Nationalfonds) as well as the German National Merit Foundation (Studienstiftung des Deutschen Volkes) is acknowledged.

(39) The too strong repulsive exponent in cvff prevents an accurate balance of bond geometry, torsion barriers, liquid density, and vaporization energy for alkanes. However, we also employed an extended cvff for our systems. The improvement for mica has no impact on the results, but the dynamics of the alkyl chains yield only a good qualitative agreement.

**Table of Content Graphic**

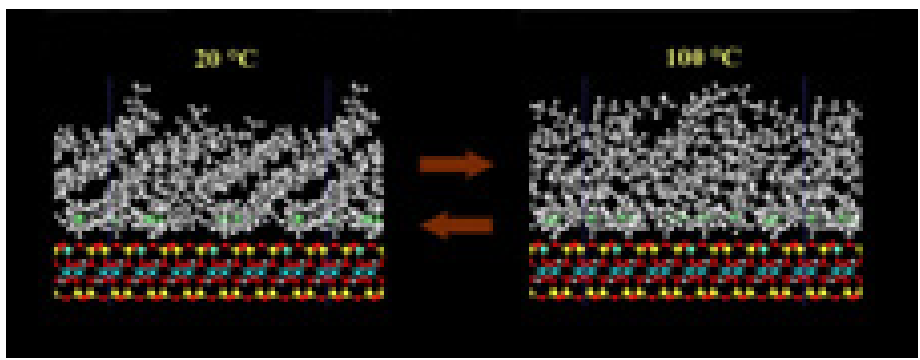